\documentclass[conference]{IEEEtran}
\IEEEoverridecommandlockouts
\usepackage{cite}
\usepackage{url}
\usepackage{amsmath,amssymb,amsfonts}
\usepackage{algorithm}
\usepackage{booktabs}
\usepackage{algorithmic}
\usepackage{graphicx}
\usepackage{textcomp}
\usepackage{xcolor}
\usepackage{tcolorbox}
\usepackage{enumitem}
\usepackage[caption=false]{subfig}
\def\BibTeX{{\rm B\kern-.05em{\sc i\kern-.025em b}\kern-.08em
    T\kern-.1667em\lower.7ex\hbox{E}\kern-.125emX}}
\usepackage{pgfplots}

\usepackage[colorlinks,urlcolor=blue,linkcolor=black,citecolor=black]{hyperref}

\pgfplotsset{compat=1.18}
\definecolor{myred}{HTML}{da3c3d}    
\definecolor{myblue}{HTML}{3584bb} 
\definecolor{softblue}{HTML}{9999ff} 
\definecolor{softgreen}{HTML}{99ff99}

\begin{document}

\title{CoDe-R: Refining Decompiler Output with LLMs via Rationale Guidance and Adaptive Inference\\
}


\author{
    \IEEEauthorblockN{Qiang Zhang\textsuperscript{1}, Zhongnian Li\textsuperscript{1,2,*}}
    \IEEEauthorblockA{
        \textsuperscript{1}School of Computer Science and Technology / School of Artificial Intelligence, China University of Mining and Technology,\\
        Xuzhou, China\\
        \textsuperscript{2}Mine Digitization Engineering Research Center of the Ministry of Education, China University of Mining and Technology,\\
        Xuzhou, China\\
        \{zqiang, zhongnianli\}@cumt.edu.cn
    }
    \thanks{\textsuperscript{*}Corresponding author.}
}


\maketitle

\begin{abstract}
Binary decompilation is a critical reverse engineering task aimed at reconstructing high-level source code from stripped executables. Although Large Language Models (LLMs) have recently shown promise, they often suffer from ``logical hallucinations'' and ``semantic misalignment'' due to the irreversible semantic loss during compilation, resulting in generated code that fails to re-execute. In this study, we propose Cognitive Decompiler Refinement with Robustness (CoDe-R), a lightweight two-stage code refinement framework. The first stage introduces Semantic Cognitive Enhancement (SCE), a Rationale-Guided Semantic Injection strategy that trains the model to recover high-level algorithmic intent alongside code. The second stage introduces a Dynamic Dual-Path Fallback (DDPF) mechanism during inference, which adaptively balances semantic recovery and syntactic stability via a hybrid verification strategy. Evaluation on the HumanEval-Decompile benchmark demonstrates that CoDe-R (using a 1.3B backbone) establishes a new State-of-the-Art (SOTA) in the lightweight regime. Notably, it is the first 1.3B model to exceed an Average Re-executability Rate of 50.00\%, significantly outperforming the baseline and effectively bridging the gap between efficient models and expert-level performance. Our code is available at \url{https://github.com/Theaoi/CoDe-R}.
\end{abstract}

\begin{IEEEkeywords}
Binary Decompilation, Large Language Models, Code Refinement, Rationale Guidance, Adaptive Inference, Re-executability
\end{IEEEkeywords}


\section{Introduction}

Decompilation, the process of reconstructing high-level source code from binary executables, is fundamental to software security, vulnerability discovery, and legacy system maintenance \cite{cifuentes1994reverse}. While traditional tools like IDA Pro \cite{ida} and Ghidra \cite{ghidra} serve as industry standards, they rely on rule-based control flow recovery. Consequently, they often yield pseudo-code cluttered with obscure pointer arithmetic \cite{balakrishnan2004analyzing} and unstructured jumps. These limitations stem from the irreversible semantic loss during compilation, where high-level syntactic structures and variable semantics are stripped away, making the output difficult for human analysts to comprehend.

Recently, Neural Decompilation has emerged as a promising paradigm. General-purpose Large Language Models (LLMs) such as CodeLlama \cite{roziere2023code} and DeepSeek-Coder \cite{guo2024deepseek} have revolutionized code generation. To align these foundations with binary recovery, Tan et al. proposed LLM4Decompile \cite{tan2024llm4decompile}, establishing a baseline for assembly-to-C translation. 
More recently, advanced methods have sought to improve performance through structural intermediaries \cite{wang2025salt4decompile,tan2025sk2decompile} or complex external relabeling pipelines \cite{feng2025ref}.

\begin{figure}[t!]
    \centering
    \includegraphics[width=1.0\linewidth]{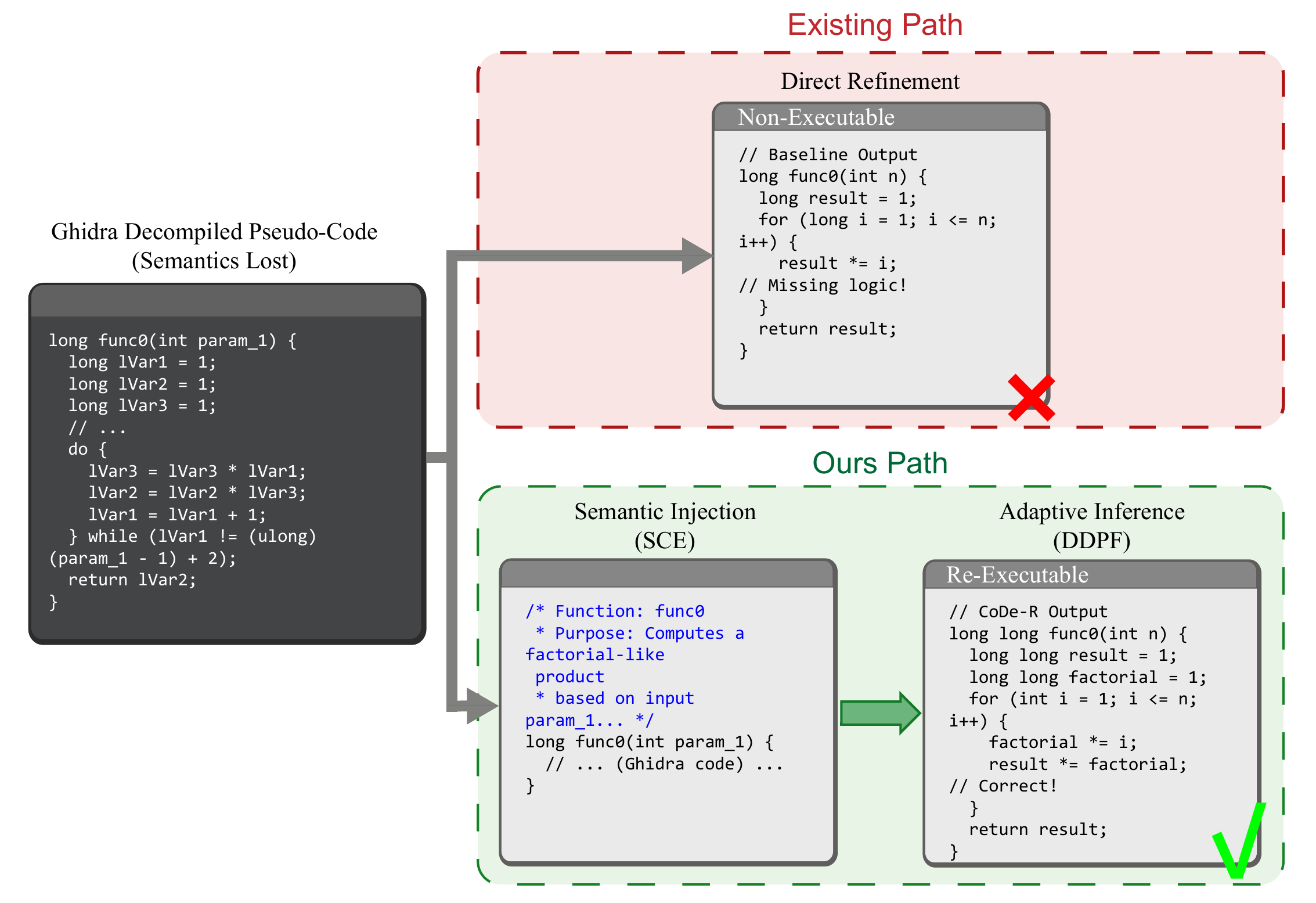}
    \caption{Comparison between existing methods and CoDe-R: While existing methods suffer from semantic loss, CoDe-R employs SCE to inject rationale, guiding the refinement of re-executable code.}
    \label{fig:teaser}
\end{figure}

However, treating decompilation as a direct translation task ($P(\text{Code}|\text{Assembly})$) presents a critical challenge. As illustrated in the ``Existing Path'' of Fig.~\ref{fig:teaser}, models often suffer from Logical Hallucination—generating code that looks syntactically correct but is functionally divergent \cite{liu2024lost}. This issue is particularly acute in lightweight models ($\approx$1.3B), which are ideal for real-time deployment but lack the deep reasoning capacity to bridge the semantic gap. Existing approaches often fail to distinguish between algorithmic intent and implementation details, leading to Semantic Misalignment where the generated code fails to re-execute \cite{ding2024semantic}.

To address this challenge, we propose Cognitive Decompiler Refinement with Robustness (CoDe-R). The name reflects our framework's goal: to act as an ``Coder'' that refines the raw output of traditional decompilers. Unlike methods that generate code from scratch, CoDe-R refines the opaque output of traditional decompilers through a cognitive process.
In the first stage (Training), we introduce Semantic Cognitive Enhancement (SCE). We adopt a Rationale-Guided Semantic Injection approach that explicitly models the intermediate reasoning process. Drawing on methodologies from Chain-of-Thought (CoT) \cite{wei2022chain}, we utilize a strong generator model to synthesize functional rationales—high-level summaries of algorithmic intent. These rationales act as Semantic Anchors, transforming the task from opaque translation into a transparent, rationale-conditional refinement process ($P(\text{Code}|\text{Input}, \text{Rationale})$).

In the second stage, recognizing that generation involves inherent uncertainty, we propose the Dynamic Dual-Path Fallback (DDPF) mechanism. Inspired by recent advances in Test-Time Compute \cite{snell2024scaling}, DDPF mitigates risk by generating two distinct candidate paths: a semantic-rich path guided by the synthesized rationale and a syntactic-robust path for stability. Crucially, we employ a hybrid verification strategy that combines compiler constraints with semantic verdicts. This allows the system to adaptively select the optimal trajectory, balancing logical recovery with execution stability.

We evaluate CoDe-R on the challenging HumanEval-Decompile benchmark \cite{tan2024llm4decompile}. Demonstrating the efficacy of our approach in resource-constrained scenarios, we implement CoDe-R using the LLM4Decompile-1.3B backbone. Experimental results show that our framework significantly outperforms the baseline across all optimization levels. Specifically, under the O0 setting, CoDe-R achieves a peak Re-executability Rate of 70.73\%, an improvement of nearly \textbf{5\%} over the baseline. These results validate that explicitly modeling "lost" functional intent allows lightweight models to punch above their weight class.

In summary, our contributions are as follows:
\begin{itemize}
    \item We propose CoDe-R, a cognitive refinement framework that enables lightweight models to master complex logic. To the best of our knowledge, this is the first work to introduce a Rationale-Guided Semantic Injection strategy in neural decompilation.
    \item We design the Dynamic Dual-Path Fallback (DDPF) mechanism. Leveraging Test-Time Compute concepts, DDPF mitigates generation uncertainty by dynamically selecting optimal trajectories via a hybrid verification strategy.
    \item We achieve an Average Re-executability Rate of \textbf{50.00\%} on HumanEval-Decompile, setting a new State-of-the-Art for lightweight neural decompilation. CoDe-R comprehensively outperforms the baseline and demonstrates robust generalization across all optimization levels.
\end{itemize}


\section{Related Work}

\subsection{End-to-End Neural Decompilation}
The paradigm of decompilation has shifted from rule-based tools like IDA Pro \cite{ida} and Ghidra \cite{ghidra} to learning-based approaches. Early attempts utilized LSTMs~\cite{katz2018using} to translate assembly into source code, while Graph Neural Networks (GNNs)~\cite{david2020neural} have been employed to predict high-level properties, such as procedure names, from stripped binaries. The emergence of LLMs has accelerated this trend. Tan et al. proposed LLM4Decompile \cite{tan2024llm4decompile}, establishing the first open-source foundation for assembly-to-C translation. 

Recent works have focused on enhancing this direct mapping ($P(\text{Code}|\text{Assembly})$) through structural intermediaries or context augmentation. 
CodeInverter \cite{liu2025codeinverter} augments the input with Control Flow Graphs (CFG) to improve structural recovery. 
To bridge the abstraction gap, Salt4Decompile \cite{wang2025salt4decompile} proposes inferring a Source-level Abstract Logic Tree (SALT) as an intermediate step, while SK2Decompile \cite{tan2025sk2decompile} introduces a ``Skeleton-to-Skin'' approach, first recovering the syntactic structure and then predicting identifiers. 
In a parallel direction, ReF Decompile \cite{feng2025ref} achieves state-of-the-art performance by integrating variable relabeling and function call graph analysis to enhance the model's understanding of data flow.
Similarly, D-LiFT \cite{zou2025dlift} utilizes Reinforcement Learning (RL) to align the decompiler backend with code quality metrics.

However, these methods typically rely on explicit structural representations or external static analysis aids. In contrast, CoDe-R prioritizes intrinsic semantic intent. By distilling functional rationales via SCE, we ensure the model captures the algorithmic logic ($z$) before synthesizing the implementation ($y$), reducing logical hallucinations without the need for complex intermediate languages or heavy pre-processing tools.

\subsection{Refinement and Neuro-Symbolic Approaches}
A parallel line of research focuses on refining the output of traditional decompilers rather than generating code from scratch. 
DeGPT~\cite{hu2024degpt} leverages LLMs to improve the readability of Ghidra-generated pseudo-code by renaming variables and simplifying control structures, while Wong et al.~\cite{wong2023refining} focus on refining decompiled code to restore recompilability. 
To enhance accuracy in variable renaming, LMPA~\cite{xu2023lmpa} proposes a neuro-symbolic synergy, using program analysis to propagate context for better prediction. 
While effective at polishing code, these refinement methods are fundamentally bound by the structural errors of the underlying traditional decompiler. 
If the initial control flow is broken (common in O3 optimization), refinement models struggle to correct the underlying logic. 
CoDe-R avoids this dependency by directly reconstructing logic from pseudo-code using a rationale-guided cognitive process.

\subsection{Augmented Generation with Rationales}
Standard LLMs struggle with complex reasoning without explicit guidance. Recent research in Chain-of-Thought (CoT) \cite{wei2022chain} and Scratchpads \cite{nye2021show} demonstrates that providing intermediate reasoning steps significantly boosts performance on complex tasks. However, applying CoT directly during inference for decompilation is computationally expensive and prone to error propagation due to the verbose nature of assembly code.

To harness this reasoning capability without the inference overhead, our SCE module adapts this insight into a Context-Augmented Generation paradigm. Instead of requiring the model to reason spontaneously at runtime, we perform Offline Rationale Generation using a strong generator to create high-quality functional summaries. These summaries are then injected as Semantic Anchors during training. This approach aligns with recent trends in Rationale-Augmented Learning \cite{hsieh2023distilling, wadhwa2024investigating}, utilizing LLM-generated reasoning to enhance small model training. Unlike refinement-based methods \cite{hu2024degpt} that polish output post-hoc, our method fundamentally alters the generation process by resolving semantic ambiguities at the input level.

\subsection{Test-Time Compute and Execution Feedback}
Recent research suggests that scaling Test-Time Compute—allocating more computational resources during inference—can be more effective than scaling model parameters \cite{snell2024scaling}. A prominent direction is Execution-based Verification, where models generate multiple candidates and select the best one based on test case execution. CodeT \cite{chen2023codet} pioneered this approach by using generated unit tests to verify code consistency. Similarly, Self-Refine \cite{madaan2024selfrefine} employs iterative feedback loops to correct errors.

The DDPF mechanism of CoDe-R draws inspiration from these strategies but is tailored for the constraints of decompilation. Instead of expensive multi-turn iterations, we employ a lightweight dual-path strategy that leverages hybrid feedback: combining the hard constraints of a compiler (re-compilability) with the soft semantic checks of a BLEU-based verifier. This allows CoDe-R to dynamically trade off between semantic fidelity and syntactic robustness without the overhead of full-scale iterative refinement.

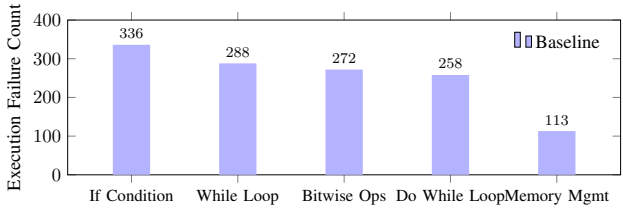
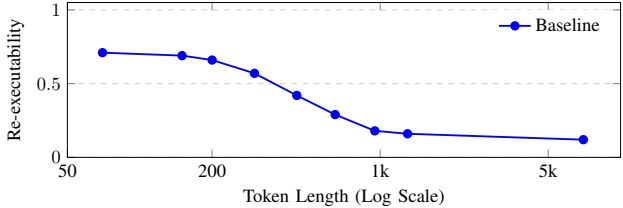
\begin{figure}[t!]
    \centering
    
    \subfloat[Top-5 Failure Count Patterns.]{
        \resizebox{0.95\linewidth}{!}{ 
        \begin{tikzpicture}
        \begin{axis}[
            ybar,
            width=12cm, height=4.5cm,
            bar width=20pt,
            enlarge x limits=0.15,
            ymin=0, ymax=400,
            ylabel={Execution Failure Count},
            symbolic x coords={If Condition, While Loop, Bitwise Ops, Do While Loop, Memory Mgmt},
            xtick=data,
            xticklabel style={rotate=0, anchor=north, font=\small},
            nodes near coords,
            nodes near coords style={font=\footnotesize, color=black, anchor=south},
            legend style={at={(0.98,0.95)}, anchor=north east, draw=none, fill=none}
        ]
        \addplot[fill=blue!30, draw=none] coordinates {
            (If Condition, 336) (While Loop, 288) (Bitwise Ops, 272) (Do While Loop, 258) (Memory Mgmt, 113)
        };
        \addlegendentry{Baseline}
        \end{axis}
        \end{tikzpicture}
        }
        \label{fig:sub_a} 
    }
    
    \vspace{0.2em} 
    
    \subfloat[Length Degradation: Performance drops as input scales.]{
        \resizebox{0.95\linewidth}{!}{
        \begin{tikzpicture}
        \begin{axis}[
            width=12cm, height=4.5cm,
            xmode=log,
            log ticks with fixed point,
            xmin=50, xmax=10000,
            ymin=0, ymax=1.05,
            xlabel={Token Length (Log Scale)},
            ylabel={Re-executability},
            xtick={50, 200, 1000, 5000},
            xticklabels={50, 200, 1k, 5k},
            ymajorgrids=true,
            grid style={dashed, gray!40},
            legend style={at={(0.98,0.95)}, anchor=north east, draw=none}
        ]
        \addplot[color=blue, mark=*, line width=1pt] coordinates {
            (70, 0.71) (150, 0.69) (200, 0.66) (300, 0.57)
            (450, 0.42) (650, 0.29) (950, 0.18) (1300, 0.16) (7000, 0.12)
        };
        \addlegendentry{Baseline}
        \end{axis}
        \end{tikzpicture}
        }
        \label{fig:sub_b}
    }
    
    \caption{Motivation Analysis (Baseline: LLM4Decompile-Ref-1.3B on HumanEval-Decompile). (a) The model struggles with control flow, indicating superficial learning \cite{ding2024semantic}. (b) Re-executability drops with length \cite{liu2024lost}.}
    \label{fig:motivation_combined}
\end{figure}


\section{Motivation and Key Insights}

We analyze the limitations of current neural decompilers to articulate the intuitions driving CoDe-R.

\subsection{Insight 1: Decompilation requires semantic guidance}
Existing methods predominantly adopt a ``Direct Mapping'' paradigm ($P(Y|X)$), mapping assembly ($X$) directly to source ($Y$). However, this is ill-posed due to the irreversible semantic loss in compilation. Compiler optimizations often map distinct source codes to identical assembly \cite{bacon1994compiler}, leaving $X$ insufficient to uniquely determine $Y$.

Fig.~\ref{fig:motivation_combined}(a) shows that errors are not uniformly distributed; the model fails most on semantic-heavy patterns (e.g., control flow), suggesting it captures superficial correlations rather than logic \cite{ding2024semantic, yakdan2015no}. Furthermore, Fig.~\ref{fig:motivation_combined}(b) reveals a performance drop as token count increases, aligning with the ``Lost-in-the-Middle'' phenomenon \cite{liu2024lost}. 
To mitigate this, we introduce Functional Rationales as domain-specific ``Semantic Landmarks'' \cite{mohtashami2023landmark}, transforming the target to $P(Y|X, Z)$ to anchor generation on \textit{what} to do before \textit{how}.

\subsection{Insight 2: The Trade-off between Rationale and Rigidity}
We observe a tension between two paradigms: Rationale-Guided Generation captures high-level intent but may violate syntax, while Direct Generation ensures robustness but misses logical dependencies. 
To resolve this, our Dynamic Dual-Path Fallback (DDPF) decouples the objectives. Inspired by Snell et al. \cite{snell2024scaling}, we leverage Test-Time Compute to generate candidates from both paradigms and use a hybrid verification strategy to dynamically select the optimal trajectory.


\begin{figure*}[t!]
    \centering
    \includegraphics[width=0.85\linewidth]{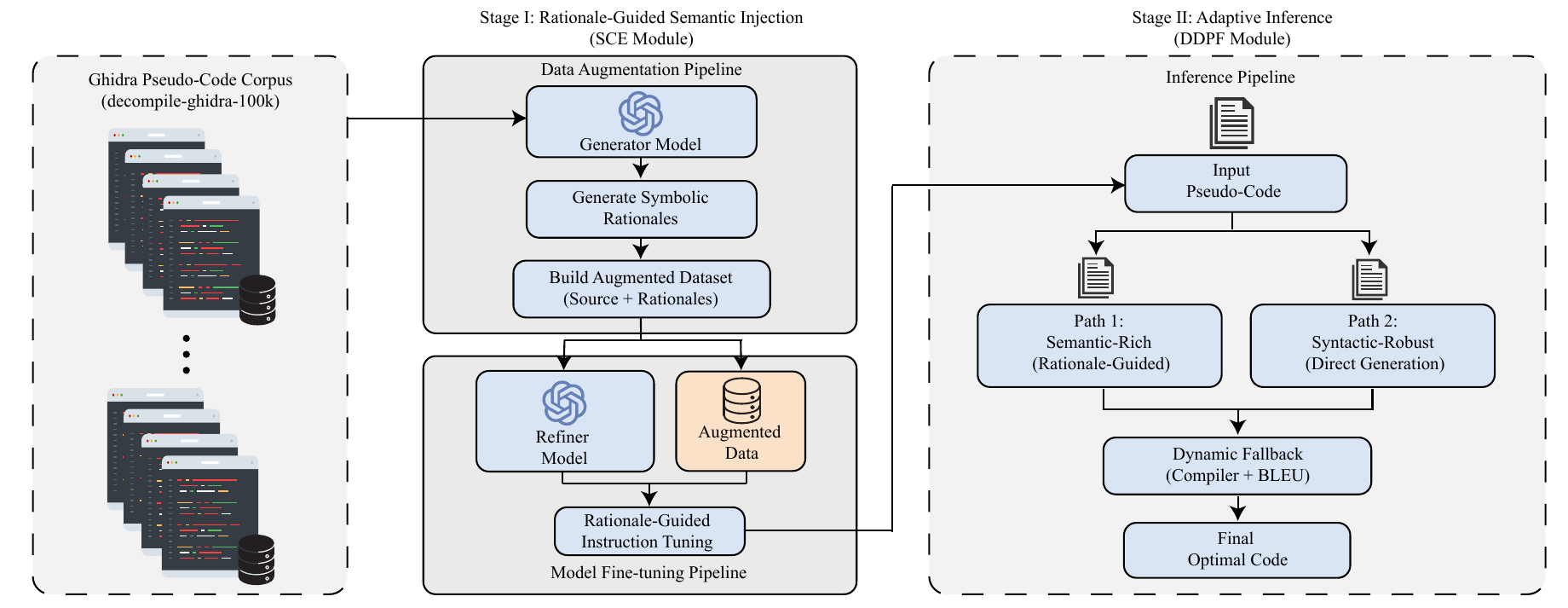} 
    \caption{The overview of CoDe-R. The framework operates in two stages: Stage I employs SCE to train the model via rationale-conditional generation; Stage II utilizes DDPF to dynamically select between semantic-rich and syntactic-robust paths via a hybrid verification strategy.}
    \label{fig:framework}
\end{figure*}

\section{Proposed Method}

We propose CoDe-R, a cognitive framework designed to optimize decompiler-generated pseudo-code. As illustrated in Fig.~\ref{fig:framework}, our pipeline operates in two main stages: Rationale-Guided Semantic Injection (SCE) during training and Adaptive Inference (DDPF) during inference.

\subsection{Stage I: Rationale-Guided Semantic Injection (SCE)}
Datasets $\mathcal{D} = \{(x_i, y_i)\}$ consist of pairs where $x_i$ is the pseudo-code decompiled from assembly via a decompiler and $y_i$ is the ground-truth source code. Direct translation $P(y|x)$ is ill-posed due to the severe semantic loss. To resolve this, we propose an Input Augmentation strategy that introduces an explicit Semantic Anchor.

We formulate the refinement task as rationale-augmented conditional generation. We posit that the generation of source code $y$ depends not only on the input pseudo-code $x$ but also on a Functional Rationale $z$, which captures the high-level algorithmic intent. Let $\theta$ denote the learnable parameters of the refinement model $\mathcal{M}*{ref}$, and let $\phi$ denote the parameters of the rationale generator $\mathcal{M}*{gen}$. In principle, the marginal generation probability can be written as:

\begin{equation}
P_\theta(y | x) = \sum_{z} P_\phi(z | x) P_\theta(y | x, z).
\end{equation}

Since exact marginalization over natural-language rationales is intractable, we approximate it with a single generated rationale. Specifically, we construct a prompt $\mathcal{P}*{gen}$ instructing $\mathcal{M}*{gen}$ to analyze the pseudo-code logic and generate a concise Symbolic Rationale $\hat{z}_i$:

\begin{equation}
\hat{z}*i = \mathcal{M}*{gen}(x_i, \mathcal{P}_{gen}),
\end{equation}

where $\hat{z}_i$ contains high-level intent descriptions. Under this approximation, the refinement objective becomes rationale-conditioned generation:

\begin{equation}
P_\theta(y_i | x_i) \approx P_\theta(y_i | x_i, \hat{z}_i).
\end{equation}

The core of SCE is to train the model $\mathcal{M}_{ref}$ to utilize the injected rationale. We employ Instruction Tuning following the standard Alpaca format \cite{alpaca}, where the input instruction explicitly includes the generated rationale $\hat{z}$. The optimization objective is to maximize the likelihood of the source code given the augmented context:

\begin{equation}
\mathcal{L}*{SCE}(\theta) = - \sum*{t=1}^{|y|} \log P_\theta(y_t | x, \hat{z}, y_{<t}).
\end{equation}
By explicitly conditioning on $z$, the model treats the rationale as semantic guidance, effectively pruning the search space and reducing generation ambiguity. From an information-theoretic perspective, the rationale acts as an information bottleneck \cite{tishby2000information} that filters implementation noise while preserving semantic intent.

\subsection{Stage II: Adaptive Inference with DDPF}
During inference, we face a dilemma: relying solely on injected rationales can be risky if the generated rationale contains noise (Hallucination Propagation), while ignoring them loses semantic depth. To balance this, we propose the Dynamic Dual-Path Fallback (DDPF) mechanism.

\begin{figure}[t!]
   \centering
   \includegraphics[width=1.0\linewidth]{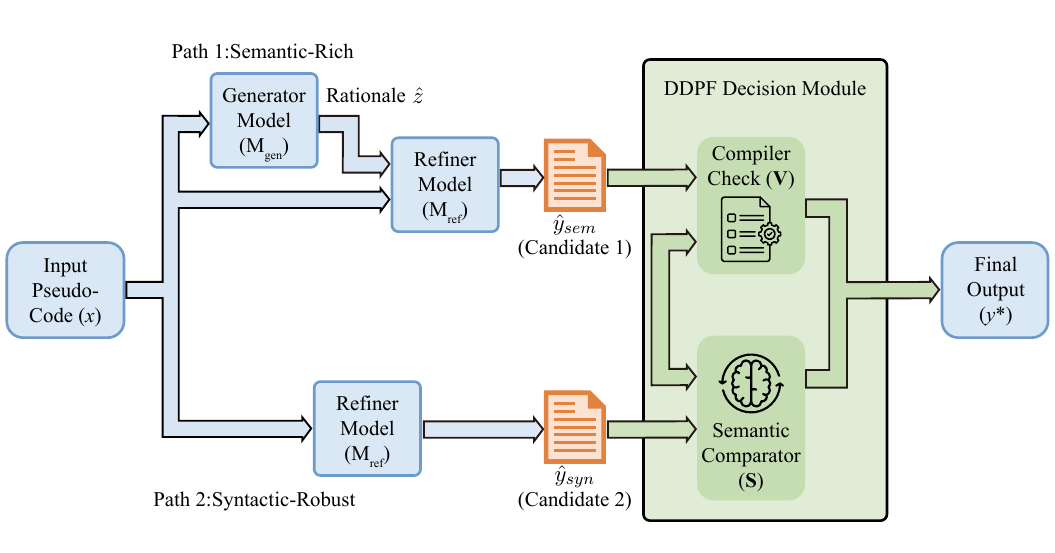} 
   \caption{The running workflow of the DDPF mechanism.}
   \label{fig:fallback}
\end{figure}

As illustrated in the inference stage of Fig.~\ref{fig:fallback}, the system maintains two parallel inference trajectories:

\subsubsection{Path 1: Semantic-Rich Generation}
This path restricts the training condition to maximize semantic recovery. 
First, we reuse the generation prompt $\mathcal{P}_{gen}$ to prompt the model to predict a functional rationale on the fly:
\begin{equation}
    \hat{z} = \mathcal{M}_{gen}(x, \mathcal{P}_{gen}). 
\end{equation}

Subsequently, the model synthesizes the source code $\hat{y}_{sem}$ utilizing this predicted rationale as a semantic anchor:
\begin{equation}
    \hat{y}_{sem} = \mathcal{M}_{ref}(x, \hat{z}). 
\end{equation}

This path excels at capturing complex algorithmic logic by grounding the generation in semantic guidance.

\subsubsection{Path 2: Syntactic-Robust Generation}
To ensure robustness when rationale generation fails, we reuse the exact same $\mathcal{M}_{ref}$ for direct generation, querying it using only the pseudo-code x:
\begin{equation}
    \hat{y}_{syn} = \mathcal{M}_{ref}(x, \emptyset). 
\end{equation}

This path forces the model to rely on its internal pattern-matching capabilities, acting as a Syntactic Stabilizer that ensures basic syntactic correctness.

\subsubsection{Hybrid Verification Strategy}
We employ a Re-Compilation Consistency strategy to select the optimal output. Since ground-truth source code is unavailable at inference, we utilize the underlying assembly as the reference.

We define the consistency score $\mathbb{S}(y)$ as the BLEU similarity between the original binary's assembly and the re-compiled assembly of the generated code. Let $\mathcal{C}(y)$ denote the compiler function that converts source code $y$ back to assembly, and $x_{asm}$ denote the original input assembly. The score is calculated as:
\begin{equation}
    \mathbb{S}(y) = \text{BLEU}(\mathcal{C}(y), x_{asm}). 
\end{equation}

Let $\mathbb{V}(y)=1$ denote that code $y$ successfully compiles. The final output $y^*$ is selected by prioritizing the semantic path ($\hat{y}_{sem}$), provided it compiles and maintains higher (or equal) assembly-level consistency than the robust path ($\hat{y}_{syn}$):

\begin{equation}
y^* = 
\begin{cases} 
\hat{y}_{sem} & \text{if } \mathbb{V}(\hat{y}_{sem}) \land (\neg\mathbb{V}(\hat{y}_{syn}) \lor \mathbb{S}(\hat{y}_{sem}) \ge \mathbb{S}(\hat{y}_{syn})). \\
\hat{y}_{syn} & \text{otherwise}. 
\end{cases}
\end{equation}

This mechanism effectively acts as a Semantic Cycle-Consistency Check. By comparing the re-compiled assembly against the original, we verify whether the generated high-level logic ($\hat{y}$) faithfully preserves the original control flow and data operations, filtering out candidates that are syntactically valid but semantically divergent.


\section{Experimental Setup}

\subsection{Datasets}
To rigorously evaluate the effectiveness of semantic injection, we utilized two distinct datasets. For training, we adopted the Decompile-Ghidra-100k dataset \cite{tan2024llm4decompile}, filtering the original 100,000 pairs down to 86,536 high-quality C/C++ source and pseudo-code pairs. For evaluation, we employed the HumanEval-Decompile benchmark \cite{tan2024llm4decompile, chen2021evaluating}, a recognized test set containing 164 samples. To simulate real-world compilation diversity, each problem was compiled under four optimization levels (O0, O1, O2, O3), resulting in a total of 656 test samples.

\begin{figure}[t!]
\vspace{-1em} 
\centering
\begin{tcolorbox}[
    colback=gray!5, 
    colframe=black!75, 
    boxrule=0.8pt, 
    left=4pt, right=4pt, top=4pt, bottom=4pt, 
    arc=0pt, outer arc=0pt 
]
\footnotesize 
\textbf{Instruction:} You are an expert in binary reverse engineering. Analyze the provided source code and summarize its high-level functionality.

\textbf{Guidelines:} (1) Use multi-line comments only at the very beginning of
the function; (2) The comment block must strictly include function name
and purpose;

\textbf{Format:} Input: \{Pseudo Code\} $\rightarrow$ Output: \{Annotated Code ($x, z$)\}
\end{tcolorbox}
\vspace{-0.5em} 
\caption{Simplified Prompt Template for Generator. We query the model to extract high-density semantic anchors ($z$) using these instructions. For the unabridged prompt incorporating expert reverse-engineering heuristics, please refer to Appendix A.}
\label{fig:prompt_template}
\vspace{-1em} 
\end{figure}

To implement our Semantic Cognitive Enhancement (SCE), we utilized the Qwen3 \cite{team2025qwen3} to synthesize Symbolic Rationales ($z$). The simplified prompting strategy is illustrated in Fig.~\ref{fig:prompt_template}. 

For training, we strictly filtered out 13,464 samples where the generator failed to yield valid comments or exceeded length limits. This produced a refined corpus of 86,536 high-quality pairs $\{(z_i \oplus x_i, y_i)\}$, adhering to \cite{gunasekar2023textbooks}, where the functional rationale $z_i$ is concatenated with pseudo-code $x_i$. Conversely, for the testing set, we retained all 656 samples to ensure a fair, 100\% coverage comparison against the baseline.

\subsection{Evaluation Metrics}
We employ three distinct metrics to comprehensively evaluate performance. The primary metric is the Re-executability Rate ($R_{\text{re-exec}}$). It measures the percentage of generated code that not only compiles successfully but also achieves the expected functionality:
\begin{equation}
    R_{\text{re-exec}} = \frac{1}{N} \sum_{i=1}^{N} \mathbb{I}(\text{Exec}(C_i, T_i)), 
\end{equation}
where $N$ is the total number of test samples, $\mathbb{I}(\cdot)$ is the indicator function which equals 1 if the condition holds and 0 otherwise, $C_i$ is the generated code, and $T_i$ represents the unit tests. Additionally, we use BLEU-4 \cite{papineni-etal-2002-bleu} to measure textual similarity with the ground truth and the Compile Rate as a baseline indicator of syntactic validity.

\subsection{Implementation Details}
We selected LLM4Decompile-Ref-1.3B \cite{tan2024llm4decompile} as our Refiner Model backbone. The model was fine-tuned using the standard causal language modeling objective, optimized via HybridAdam with a learning rate of $2\times 10^{-6}$ and a cosine decay scheduler. We set the micro-batch size to 8 per device and trained for 2 epochs with a maximum sequence length of 2048 tokens. 
Experiments were conducted on a heterogeneous computing cluster: Refiner Model training was performed on a node with 4 $\times$ RTX 4090D. For inference, the model runs on a single RTX 4090D, while the Generator's rationale generation is offloaded to a node equipped with a single H20-NVLink.


\begin{table}[t!]
\caption{Comparison of Re-executability Rate (\%) on HumanEval-Decompile (Focus on lightweight methods)}
\begin{center}
\resizebox{\linewidth}{!}{
\begin{tabular}{lccccc}
\toprule
{\bfseries Method} & {\bfseries O0} & {\bfseries O1} & {\bfseries O2} & {\bfseries O3} & {\bfseries Avg} \\
\midrule
\multicolumn{6}{l}{{\itshape Base Tool}} \\
Ghidra (Base) & 33.54 & 16.46 & 15.85 & 13.41 & 19.82 \\
\midrule
\multicolumn{6}{l}{{\itshape Refinement Methods}} \\
+Idioms & \underline{70.73} & 27.44 & 13.41 & 12.20 & 30.95 \\
+LLM4Decompile-Ref (1.3B)* & 65.85 & 36.59 & 40.24 & 36.59 & 44.82 \\
{\bfseries +Ours (CoDe-R)} & \underline{70.73} & {\bfseries 46.34} & {\bfseries 42.07} & {\bfseries 40.85} & {\bfseries 50.00} \\
\midrule
\multicolumn{6}{l}{{\itshape End to End Method}} \\
Nova-1.3B \cite{jiang2023nova} & 37.53 & 21.71 & 22.68 & 18.75 & 25.17 \\
Nova-6.7B \cite{jiang2023nova} & 48.78 & 30.58 & 30.85 & 27.23 & 34.36 \\
CodeInverter (1.3B)  \cite{liu2025codeinverter} & {\bfseries 71.34} & 39.63 & \underline{42.07} & 40.24 & \underline{48.32} \\
\midrule
\multicolumn{6}{l}{{\itshape General-purpose LLMs}} \\
Qwen-Plus & 20.12 & 7.93 & 5.49 & 8.54 & 10.52 \\
GPT-4o & 34.15 & 11.59 & 15.24 & 10.37 & 17.84 \\
DeepSeek-V3 & 67.07 & 37.20 & 37.80 & 37.20 & 44.82\\
\bottomrule
\multicolumn{6}{l}{\footnotesize * Indicates the baseline model. {\bfseries Bold}: Best. \underline{Underline}: Second Best.}
\end{tabular}
}
\label{tab:main_results}
\end{center}
\end{table}

\section{Results and Discussion}

\subsection{Main Results}
Table~\ref{tab:main_results} compares CoDe-R against methods across varying parameter scales.
CoDe-R achieves the highest average re-executability (50.00\%), outperforming the baseline by \textbf{5.18\%}.
Compared to the structure-aware CodeInverter \cite{liu2025codeinverter}, CoDe-R shows superior robustness at higher optimization levels (O1-O3), proving the efficacy of semantic injection for complex logic.

Crucially, CoDe-R demonstrates exceptional parameter efficiency. It not only doubles the performance of Nova-1.3B (25.17\%) but also significantly surpasses the larger Nova-6.7B (34.36\%) \cite{jiang2023nova}.
This result highlights that domain-specific cognitive alignment (via SCE and DDPF) is a more effective driver of performance than mere parameter scaling.
Furthermore, CoDe-R outperforms massive generalist models like DeepSeek-V3 (44.82\%) and GPT-4o (17.84\%), solidifying its position as the state-of-the-art in lightweight neural decompilation refinement.

\subsection{Error Pattern Analysis}
Recall the motivation analysis in Section III (Fig.~\ref{fig:motivation_combined}(a)), where we identified that existing methods suffer from severe ``Deep Program Semantics'' deficits \cite{ding2024semantic}. To verify whether CoDe-R effectively addresses this issue, we conducted a fine-grained comparison on the top-5 failure count patterns.

\begin{figure}[t!]
    \centering
    \begin{tikzpicture}
    \begin{axis}[
        ybar,
        width=0.95\linewidth, height=6.5cm,
        bar width=12pt,
        enlarge x limits=0.15,
        ymin=45, ymax=68,
        ylabel={Execution Failure Rate (\%)},
        symbolic x coords={If Condition, While Loop, Bitwise Ops, Do While Loop, Memory Mgmt},
        xtick=data,
        xticklabel style={rotate=25, anchor=north east, align=right, font=\rmfamily\small},
        yticklabel style={font=\rmfamily},
        ymajorgrids=true,
        grid style={dashed, gray!30},
        nodes near coords,
        nodes near coords style={font=\tiny, color=black, yshift=2pt},
        legend style={
            at={(0.5,-0.35)},
            anchor=north,
            legend columns=-1,
            draw=black,            
            fill=none,
            font=\rmfamily
        }
    ]
    \addplot[fill=softblue, draw=none] coordinates {
        (If Condition, 54.8)
        (While Loop, 57.6)
        (Bitwise Ops, 60.7)
        (Do While Loop, 58.9)
        (Memory Mgmt, 61.4)
    };
    \addlegendentry{Baseline}

    \addplot[fill=softgreen, draw=none] coordinates {
        (If Condition, 52.9)
        (While Loop, 56.8)
        (Bitwise Ops, 60.5)
        (Do While Loop, 57.8)
        (Memory Mgmt, 59.8)
    };
    \addlegendentry{CoDe-R}
    \end{axis}
    \end{tikzpicture}
    \vspace{0.5em}
    \caption{Failure Rate Comparison on Top-5 Failure Count Patterns: CoDe-R (Green) consistently reduces failure rates compared to Baseline (Blue). Note the significant drop in if-condition and  memory-mgmt.}
    \label{fig:error_analysis}
\end{figure}
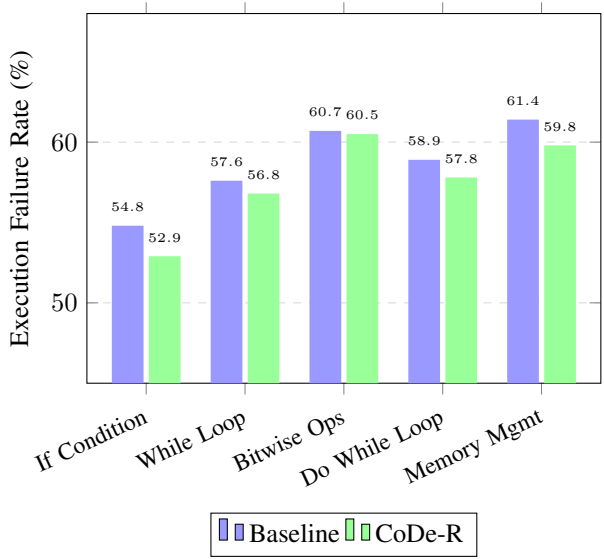

As shown in Fig.~\ref{fig:error_analysis}, CoDe-R consistently reduces failure rates, yet the magnitude of improvement varies non-uniformly, aligning with established program comprehension theories. We observe the most distinct reductions in if\_condition and memory\_management. As established in classic reverse engineering literature \cite{yakdan2015no}, recovering structured control flow and variable abstractions represents the primary ``semantic gap.'' The substantial gains here confirm that our Symbolic Rationale effectively provides the missing functional intent, enabling the model to reconstruct logic that relies on global understanding rather than local syntax.

Conversely, the improvement in bitwise\_ops is minimal. Bitwise operations often stem from compiler optimizations (e.g., strength reduction) or low-level arithmetic \cite{bacon1994compiler}, which serve as local implementation details rather than high-level algorithmic intent. Since these patterns rely more on local syntax than global semantics, the baseline model captures them sufficiently, and the high-level rationale offers limited additional guidance.

This differential improvement strongly supports our core hypothesis: CoDe-R effectively restores the semantic information lost during compilation, thereby significantly enhancing the re-executability of code patterns with high semantic demands.

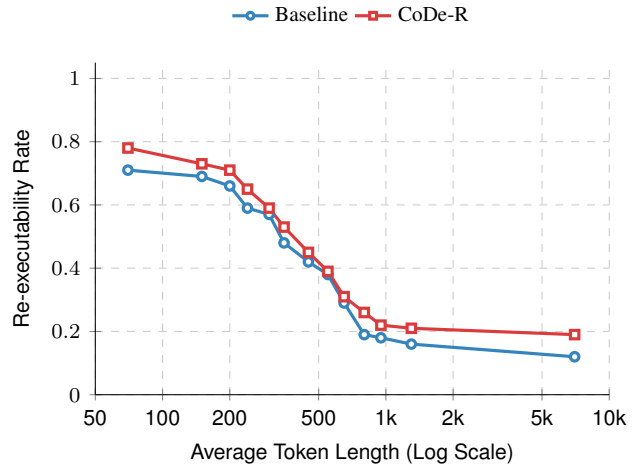
\begin{figure}[t!]
    \centering
    \begin{tikzpicture}
    \begin{axis}[
        width=0.95\linewidth, height=6cm,
        xmode=log,
        log ticks with fixed point,
        label style={font=\sffamily\footnotesize},
        tick label style={font=\sffamily\footnotesize},
        xmin=50, xmax=10000,
        ymin=0, ymax=1.05,
        xlabel={Average Token Length (Log Scale)},
        ylabel={Re-executability Rate},
        xtick={50, 100, 200, 500, 1000, 2000, 5000, 10000},
        xticklabels={50, 100, 200, 500, 1k, 2k, 5k, 10k},
        ytick={0, 0.2, 0.4, 0.6, 0.8, 1.0},
        xmajorgrids=true,
        ymajorgrids=true,
        grid style={dashed, gray!40},
        axis lines=left,
        axis line style={-},
        legend style={at={(0.5, 1.08)}, anchor=south, legend columns=-1, draw=none, fill=none, font=\rmfamily\footnotesize},
        legend image post style={scale=0.8}
    ]
    \addplot[color=myblue, mark=*, mark options={fill=white, scale=0.8}, line width=1.2pt]
    coordinates {
        (70, 0.71) (150, 0.69) (200, 0.66) (240, 0.59) (300, 0.57)
        (350, 0.48) (450, 0.42) (550, 0.38) (650, 0.29) (800, 0.19)
        (950, 0.18) (1300, 0.16) (7000, 0.12)
    };
    \addlegendentry{Baseline}
    \addplot[color=myred, mark=square*, mark options={fill=white, scale=0.8}, line width=1.2pt]
    coordinates {
        (70, 0.78) (150, 0.73) (200, 0.71) (240, 0.65) (300, 0.59)
        (350, 0.53) (450, 0.45) (550, 0.39) (650, 0.31) (800, 0.26)
        (950, 0.22) (1300, 0.21) (7000, 0.19)
    };
    \addlegendentry{CoDe-R}
    \end{axis}
    \end{tikzpicture}
    \caption{Re-executability Rate vs. Code Length: CoDe-R (Red) demonstrates superior robustness in the long-context regime ($>1000$ tokens) compared to Baseline (Blue), confirming the anchoring effect of rationales.}
    \label{fig:length_analysis}
\end{figure}

\subsection{Impact of Code Complexity}

To further validate the efficacy of our SCE module in mitigating the ``Lost-in-the-Middle'' challenge discussed in Section III, we analyzed the correlation between code length and re-executability. Fig.~\ref{fig:length_analysis} shows that CoDe-R (Red Line) demonstrates consistent superiority over the Baseline (Blue Line). The underlying mechanism varies across complexity regimes, which we interpret through the lens of Information Theory.

In the regime of Short Contexts ($<300$ tokens), CoDe-R achieves maximal gains. As noted by Ding et al. \cite{ding2024semantic}, short functions are often dominated by a single algorithmic intent; here, our generated rationale ($z$) provides near-perfect semantic coverage, bridging the gap between assembly and source code with high precision. However, as complexity increases to Medium Contexts ($400-800$ tokens), the performance gap narrows slightly. We attribute this to the Information Bottleneck principle \cite{tishby2000information}: medium-length code often resides in a ``complexity valley'', which is complex enough to require specific implementation details yet short enough for the baseline to memorize local patterns.

Crucially, the divergence becomes most pronounced in Long Contexts ($>1000$ tokens). Consistent with our motivation, the baseline suffers from a catastrophic drop in the long tail due to context drift. In contrast, CoDe-R maintains a significant margin. This confirms that our Symbolic Rationale effectively functions as a ``Semantic Landmark'' \cite{mohtashami2023landmark}, allowing the model to maintain logical coherence even when the local context window is saturated.

\subsection{Ablation Study}
To systematically evaluate the contribution of each design component in CoDe-R, we conducted comprehensive ablation studies. We dissect the framework to analyze four critical aspects: the isolated efficacy of the Semantic Cognitive Enhancement (SCE) module, the architectural necessity of the Dynamic Dual-Path Fallback (DDPF) mechanism, the impact of rationale granularity, and the optimal injection strategy.

\subsubsection{Effect of SCE Module}
The core premise of CoDe-R is that explicitly injecting functional rationales serves as a critical semantic anchor. To isolate this effect, we examine the performance of Path 1 Only, which represents the model operating purely in the Rationale-Guided mode (trained with SCE).

To isolate the efficacy of explicit semantic injection, we compare Path 1 Only (Semantic-Rich) directly with Path 2 Only (Syntactic-Robust). As shown in Table~\ref{tab:ablation_ddpf}, Path 1 achieves a higher average re-executability (48.32\%) compared to Path 2 (47.71\%). This result validates the hypothesis of Rationale-Guided Generation: even when the model is capable of direct synthesis, explicitly conditioning the process on the functional summaries ($z$) further constrains the search space. This proves that the injected rationale serves as a necessary Semantic Anchor, bridging the gap between implicit intent and explicit implementation effectively.

\begin{table}[t!]
\caption{Ablation Study of Component Contributions}
\begin{center}
\resizebox{\linewidth}{!}{
\begin{tabular}{lccccc}
\toprule
\textbf{Configuration} & \textbf{O0} & \textbf{O1} & \textbf{O2} & \textbf{O3} & \textbf{Avg} \\
\midrule
Path 2 Only {\itshape (Syntactic-Robust)} & 67.68 & 42.07 & \bfseries 42.68 & 38.41 & 47.71 \\
Path 1 Only {\itshape (Semantic-Rich)} & \bfseries 70.73 & 42.68 & \bfseries 42.68 & 37.20 & 48.32 \\
\midrule
\bfseries CoDe-R (DDPF Combined) & \bfseries 70.73 & \bfseries 46.34 & 42.07 & \bfseries 40.85 & \bfseries 50.00 \\
\bottomrule
\end{tabular}
}
\label{tab:ablation_ddpf}
\end{center}
\end{table}

\subsubsection{Effect of DDPF Mechanism}
To validate the dual-path design, we compare CoDe-R against individual paths in Table~\ref{tab:ablation_ddpf}. Results show a complementary relationship: Path 1 excels in structure-preserving scenarios (O0: 70.73\%), while Path 2 demonstrates resilience in optimized settings (O3). CoDe-R computes the union of these strategies, achieving the highest average re-executability of 50.00\%.

\begin{table}[t!]
\caption{Compilability Analysis (Pass@1-Compile \%)}
\begin{center}
\resizebox{\linewidth}{!}{
\begin{tabular}{lccccc}
\toprule
\textbf{Model} & \textbf{O0} & \textbf{O1} & \textbf{O2} & \textbf{O3} & \textbf{Avg} \\
\midrule
Baseline & 89.63 & 88.41 & \bfseries 93.29 & 89.63 & 90.24 \\
Path 1 {\itshape (Semantic)} & 85.37 & 87.80 & 82.93 & 79.88 & 83.99 \\
Path 2 {\itshape (Robust)} & \bfseries 90.24 & 87.80 & 90.85 & 89.02 & 89.48 \\
\bfseries CoDe-R & \bfseries 90.24 & \bfseries 91.46 & 90.24 & \bfseries 90.85 & \bfseries 90.70 \\
\bottomrule
\end{tabular}
}
\label{tab:compilability}
\end{center}
\end{table}

To further verify the mechanism, Table~\ref{tab:compilability} analyzes syntactic validity. While Path 1 suffers from lower compilability (Avg: 83.99\%) due to aggressive reasoning, the robust Path 2 (Avg: 89.48\%) acts as a Syntactic Stabilizer. DDPF effectively leverages this to fix syntax errors, boosting the overall compilability to 90.70\%.

\subsubsection{Effect of Rationale Granularity}
To determine our design choice for Rationale Granularity, we compared our Concise Rationale strategy (strictly Function Name and Purpose) against a Detailed Rationale strategy (expanded with Inputs, Outputs, and Implicit Operations). As shown in Table~\ref{tab:granularity}, the Concise strategy outperforms the Detailed approach on average (47.56\% vs. 46.80\%). We attribute the performance drop in the Detailed setting to a poor Signal-to-Noise Ratio. A verbose rationale consumes context window and introduces speculative fields prone to hallucination. These hallucinations act as semantic noise, distracting the model rather than guiding it. Thus, prioritizing Semantic Density over volume proves critical for robust refinement.

\begin{table}[t!]
\caption{Impact of Rationale Granularity on Re-executability (\%)}
\begin{center}
\resizebox{\linewidth}{!}{
\begin{tabular}{lccccc}
\toprule
\textbf{Configuration} & \textbf{O0} & \textbf{O1} & \textbf{O2} & \textbf{O3} & \textbf{Avg} \\
\midrule
Detailed & 67.07 & 39.63 & \bfseries 43.90 & 36.59 & 46.80 \\
\bfseries Concise & \bfseries 70.73 & \bfseries 42.07 & 39.63 & \bfseries 37.80 & \bfseries 47.56 \\
\bottomrule
\end{tabular}
}
\label{tab:granularity}
\end{center}
\end{table}

\subsubsection{Effect of Injection Strategy}
We further investigated whether the model benefits more from utilizing reasoning or learning to reason. We compared a Source-Only strategy , where the rationale $z$ acts solely as input context ($P(y | x, z)$), against a Full Distillation strategy that forces the model to generate the rationale before the code ($P(z, y | x, z)$). As summarized in Table~\ref{tab:injection_strategy}, Source-Only consistently outperforms Distillation, particularly in the O3 setting. This aligns with findings on CoT-augmented distillation \cite{wadhwa2024investigating}: the dual objective in Distillation burdens the model, where imperfect rationale generation propagates errors to the code. By treating the rationale as a fixed Semantic Anchor, we effectively offload the reasoning burden to the Generator, allowing the model to focus entirely on translation.

\begin{table}[t!]
\caption{Comparison of Injection Strategies: Utilizing vs. Distillation (\%)}
\begin{center}
\resizebox{\linewidth}{!}{
\begin{tabular}{lccccc}
\toprule
\textbf{Injection Strategy} & \textbf{O0} & \textbf{O1} & \textbf{O2} & \textbf{O3} & \textbf{Avg} \\
\midrule
Both {\itshape (Full Distillation)} & 67.68 & \bfseries 43.90 & 37.20 & 32.93 & 45.43 \\
\bfseries Source-Only {\itshape (Utilizing)} & \bfseries 70.73 & 42.07 & \bfseries 39.63 & \bfseries 37.80 & \bfseries 47.56 \\
\bottomrule
\end{tabular}
}
\label{tab:injection_strategy}
\end{center}
\end{table}


\section{Limitations}

Despite the promising performance of CoDe-R, several limitations remain. 
First, the DDPF mechanism introduces inference overhead, increasing latency compared to single-pass methods. However, this trade-off is highly practical given the substantial 5\% gain in re-executability.
Second, our evaluation is restricted to C/C++ compiled via GCC. The generalizability of our rationale-guided approach to other compiled languages (such as Rust or Go) and diverse compiler toolchains (e.g., MSVC or Clang) requires further verification.


\section{Conclusion}

This study introduces CoDe-R, a cognitive refinement framework designed to refine decompiler-generated pseudo-code into high-quality source code. By robustly injecting semantics to adaptively enhance context via Semantic Cognitive Enhancement (SCE) and applying a Dynamic Dual-Path Fallback (DDPF) mechanism, we addressed both the logical hallucinations and semantic misalignment inherent in existing direct-mapping paradigms. Experimental results on the HumanEval-Decompile benchmark demonstrate that CoDe-R sets a new State-of-the-Art for lightweight models with an average re-executability rate of 50.00\%. This validates that explicitly recovering lost semantic information allows efficient models to punch above their weight class. This rationale-guided paradigm could be generalized to other reverse engineering tasks, indicating the potential for broader applicability in cognitive code understanding.

Future work will focus on reducing the inference latency of the dual-path mechanism through parallel decoding. Additionally, we plan to extend CoDe-R to support modern compiled languages such as Rust and Go.

\section*{Acknowledgment}

This work was supported by the National Natural Science Foundation of China (No.62306320, 61976217) and the Natural Science Foundation of Jiangsu Province (No. BK20231063).

\bibliographystyle{IEEEtran} 
\bibliography{references}    

@article{tan2024llm4decompile,
  title={Llm4decompile: Decompiling binary code with large language models},
  author={Tan, Hanzhuo and Luo, Qi and Li, Jing and Zhang, Yuqun},
  journal={arXiv preprint arXiv:2403.05286},
  year={2024}
}

@article{snell2024scaling,
  title={Scaling llm test-time compute optimally can be more effective than scaling model parameters},
  author={Snell, Charlie and Lee, Jaehoon and Xu, Kelvin and Kumar, Aviral},
  journal={arXiv preprint arXiv:2408.03314},
  year={2024}
}

@misc{ida,
  author = {{Hex-Rays}},
  title = {IDA Pro: a cross-platform multi-processor disassembler and debugger},
  year = {2024},
  howpublished = {\url{https://hex-rays.com/ida-pro/}}
}

@misc{ghidra,
  title = {Ghidra},
  year = {2023},
  howpublished = {\url{https://github.com/NationalSecurityAgency/ghidra}}
}

@inproceedings{katz2018using,
  title={Using recurrent neural networks for decompilation},
  author={Katz, Deborah S and Ruchti, Jason and Schulte, Eric},
  booktitle={2018 IEEE 25th international conference on software analysis, evolution and reengineering (SANER)},
  pages={346--356},
  year={2018},
  organization={IEEE}
}

@article{roziere2023code,
  title={Code llama: Open foundation models for code},
  author={Roziere, Baptiste and Gehring, Jonas and Gloeckle, Fabian and Sootla, Sten and Gat, Itai and Tan, Xiaoqing Ellen and Adi, Yossi and Liu, Jingyu and Sauvestre, Romain and Remez, Tal and others},
  journal={arXiv preprint arXiv:2308.12950},
  year={2023}
}

@inproceedings{wei2022chain,
  title={Chain-of-thought prompting elicits reasoning in large language models},
  author={Wei, Jason and Wang, Xuezhi and Schuurmans, Dale and Bosma, Maarten and Xia, Fei and Chi, Ed and Le, Quoc V and Zhou, Denny and others},
  booktitle={Advances in Neural Information Processing Systems},
  volume={35},
  pages={24824--24837},
  year={2022}
}

@article{nye2021show,
  title={Show your work: Scratchpads for intermediate computation with language models},
  author={Nye, Maxwell and Andreassen, Anders Johan and Gur-Ari, Guy and Michalewski, Henryk and Austin, Jacob and Bieber, David and Dohan, David and Lewkowycz, Aitor and Bosma, Maarten and Luan, David and others},
  journal={arXiv preprint arXiv:2112.00114},
  year={2021}
}

@article{gunasekar2023textbooks,
  title={Textbooks are all you need},
  author={Gunasekar, Suriya and Zhang, Yi and Aneja, Jyoti and Mendes, Caio C{\'e}sar Teodoro and Del Giorno, Allie and Gopi, Sivakanth and Javaheripi, Mojan and Kauffmann, Piero and de Rosa, Gustavo and Saarikivi, Olli and others},
  journal={arXiv preprint arXiv:2306.11644},
  year={2023}
}

@misc{alpaca,
  title={Stanford alpaca: An instruction-following llama model},
  author={Taori, Rohan and Gulrajani, Ishaan and Zhang, Tianyi and Dubois, Yann and Li, Xuechen and Guestrin, Carlos and Liang, Percy and Hashimoto, Tatsunori B},
  year={2023},
  publisher={Stanford, CA, USA}
}

@article{liu2025codeinverter,
  title={The CodeInverter Suite: Control-Flow and Data-Mapping Augmented Binary Decompilation with LLMs},
  author={Liu, Peipei and Sun, Jian and Sun, Rongkang and Chen, Li and Yan, Zhaoteng and Zhang, Peizheng and Sun, Dapeng and Wang, Dawei and Zhang, Xiaoling and Li, Dan},
  journal={arXiv preprint arXiv:2503.07215},
  year={2025}
}

@inproceedings{ding2024semantic,
  title={Semantic-aware Source Code Modeling},
  author={Ding, Yangruibo},
  booktitle={Proceedings of the 39th IEEE/ACM International Conference on Automated Software Engineering},
  pages={2494--2497},
  year={2024}
}

@book{cifuentes1994reverse,
  title={Reverse compilation techniques},
  author={Cifuentes, Cristina},
  year={1994},
  publisher={Queensland University of Technology, Brisbane}
}

@inproceedings{yakdan2015no,
  title={No More Gotos: Decompilation Using Pattern-Independent Control-Flow Structuring and Semantic-Preserving Transformations.},
  author={Yakdan, Khaled and Eschweiler, Sebastian and Gerhards-Padilla, Elmar and Smith, Matthew},
  booktitle={NDSS},
  year={2015}
}

@inproceedings{balakrishnan2004analyzing,
  title={Analyzing memory accesses in x86 executables},
  author={Balakrishnan, Gogul and Reps, Thomas},
  booktitle={International conference on compiler construction},
  pages={5--23},
  year={2004},
  organization={Springer}
}

@article{bacon1994compiler,
  title={Compiler transformations for high-performance computing},
  author={Bacon, David F and Graham, Susan L and Sharp, Oliver J},
  journal={ACM Computing Surveys (CSUR)},
  volume={26},
  number={4},
  pages={345--420},
  year={1994},
  publisher={ACM New York, NY, USA}
}

@article{liu2024lost,
  title={Lost in the middle: How language models use long contexts},
  author={Liu, Nelson F and Lin, Kevin and Hewitt, John and Paranjape, Ashwin and Bevilacqua, Michele and Petroni, Fabio and Liang, Percy},
  journal={Transactions of the association for computational linguistics},
  volume={12},
  pages={157--173},
  year={2024}
}

@article{tishby2000information,
  title={The information bottleneck method},
  author={Tishby, Naftali and Pereira, Fernando C and Bialek, William},
  journal={arXiv preprint physics/0004057},
  year={2000}
}

@article{mohtashami2023landmark,
  title={Landmark attention: Random-access infinite context length for transformers},
  author={Mohtashami, Amirkeivan and Jaggi, Martin},
  journal={arXiv preprint arXiv:2305.16300},
  year={2023}
}

@inproceedings{chen2023codet,
  title={CodeT: Code Generation with Generated Tests},
  author={Chen, Bei and Zhang, Fengji and Nguyen, Anh and Da, Zan and Bowman, Samuel R and others},
  booktitle={ICLR},
  year={2023}
}

@inproceedings{madaan2024selfrefine,
  title={Self-refine: Iterative refinement with self-feedback},
  author={Madaan, Aman and Tandon, Niket and Gupta, Prakhar and Hallinan, Skyler and Gao, Luyu and Wiegreffe, Sarah and Alon, Uri and Dziri, Nouha and Prabhumoye, Shrimai and Yang, Yiming and others},
  booktitle={Advances in Neural Information Processing Systems},
  volume={36},
  pages={46534--46594},
  year={2023}
}

@article{guo2024deepseek,
  title={DeepSeek-Coder: When the Large Language Model Meets Programming--The Rise of Code Intelligence},
  author={Guo, Daya and Zhu, Qihao and Yang, Dejian and Xie, Zhenda and Dong, Kai and Zhang, Wentao and Chen, Guanting and Bi, Xiao and Wu, Yu and Li, YK and others},
  journal={arXiv preprint arXiv:2401.14196},
  year={2024}
}

@article{chen2021evaluating,
  title={Evaluating large language models trained on code},
  author={Chen, Mark and Tworek, Jerry and Jun, Heewoo and Yuan, Qiming and Pinto, Henrique Ponde de Oliveira and Kaplan, Jared and Edwards, Harri and Burda, Yuri and Joseph, Nicholas and Brockman, Greg and others},
  journal={arXiv preprint arXiv:2107.03374},
  year={2021}
}

@article{david2020neural,
  title={Neural reverse engineering of stripped binaries using augmented control flow graphs},
  author={David, Yaniv and Alon, Uri and Yahav, Eran},
  journal={Proceedings of the ACM on Programming Languages},
  volume={4},
  number={OOPSLA},
  pages={1--28},
  year={2020},
  publisher={ACM New York, NY, USA}
}

@article{tan2025sk2decompile,
  title={SK2Decompile: LLM-based Two-Phase Binary Decompilation from Skeleton to Skin},
  author={Tan, Hanzhuo and Li, Weihao and Tian, Xiaolong and Wang, Siyi and Liu, Jiaming and Li, Jing and Zhang, Yuqun},
  journal={arXiv preprint arXiv:2509.22114},
  year={2025}
}

@article{zou2025dlift,
  title={D-LiFT: Improving LLM-based Decompiler Backend via Code Quality-driven Fine-tuning},
  author={Zou, Muqi and Cai, Hongyu and Wu, Hongwei and Basque, Zion Leonahenahe and Khan, Arslan and Celik, Berkay and Bianchi, Antonio and Xu, Dongyan and others},
  journal={arXiv preprint arXiv:2506.10125},
  year={2025}
}

@inproceedings{hu2024degpt,
  title={DeGPT: Optimizing decompiler output with LLM},
  author={Hu, Pei and Liang, Rui and Chen, Kai},
  booktitle={Network and Distributed System Security Symposium},
  year={2024}
}

@article{wong2023refining,
  title={Refining decompiled c code with large language models},
  author={Wong, Wai Kin and Wang, Huaijin and Li, Zongjie and Liu, Zhibo and Wang, Shuai and Tang, Qiyi and Nie, Sen and Wu, Shi},
  journal={arXiv preprint arXiv:2310.06530},
  year={2023}
}

@article{xu2023lmpa,
  title={Lmpa: Improving decompilation by synergy of large language model and program analysis},
  author={Xu, Xiangzhe and Zhang, Zhuo and Feng, Shiwei and Ye, Yapeng and Su, Zian and Jiang, Nan and Cheng, Siyuan and Tan, Lin and Zhang, Xiangyu},
  journal={arXiv preprint arXiv:2306.02546},
  year={2023}
}

@article{wadhwa2024investigating,
  title={Investigating mysteries of cot-augmented distillation},
  author={Wadhwa, Somin and Amir, Silvio and Wallace, Byron C},
  journal={arXiv preprint arXiv:2406.14511},
  year={2024}
}

@inproceedings{hsieh2023distilling,
  title={Distilling step-by-step! outperforming larger language models with less training data and smaller model sizes},
  author={Hsieh, Cheng-Yu and Li, Chun-Liang and Yeh, Chih-Kuan and Nakhost, Hootan and Fujii, Yasuhisa and Ratner, Alex and Krishna, Ranjay and Lee, Chen-Yu and Pfister, Tomas},
  booktitle={Findings of the Association for Computational Linguistics: ACL 2023},
  pages={8003--8017},
  year={2023}
}

@misc{team2025qwen3,
  title={Qwen3 technical report},
  author={Yang, An and Li, Anfeng and Yang, Baosong and Zhang, Beichen and Hui, Binyuan and Zheng, Bo and Yu, Bowen and Gao, Chang and Huang, Chengen and Lv, Chenxu and others},
  journal={arXiv preprint arXiv:2505.09388},
  year={2025}
}

@inproceedings{papineni-etal-2002-bleu,
  title={Bleu: a method for automatic evaluation of machine translation},
  author={Papineni, Kishore and Roukos, Salim and Ward, Todd and Zhu, Wei-Jing},
  booktitle={Proceedings of the 40th annual meeting of the Association for Computational Linguistics},
  pages={311--318},
  year={2002}
}

@article{feng2025ref,
  title={ReF Decompile: Relabeling and Function Call Enhanced Decompile},
  author={Feng, Yunlong and Li, Bohan and Shi, Xiaoming and Zhu, Qingfu and Che, Wanxiang},
  journal={arXiv preprint arXiv:2502.12221},
  year={2025}
}

@article{wang2025salt4decompile,
  title={Salt4decompile: Inferring source-level abstract logic tree for llm-based binary decompilation},
  author={Wang, Yongpan and Xu, Xin and Zhu, Xiaojie and Gu, Xiaodong and Shen, Beijun},
  journal={arXiv preprint arXiv:2509.14646},
  year={2025}
}

@article{jiang2023nova,
  title={Nova: Generative language models for assembly code with hierarchical attention and contrastive learning},
  author={Jiang, Nan and Wang, Chengxiao and Liu, Kevin and Xu, Xiangzhe and Tan, Lin and Zhang, Xiangyu and Babkin, Petr},
  journal={arXiv preprint arXiv:2311.13721},
  year={2023}
}

\onecolumn 
\appendices 

\section{Detailed Prompt for Rationale Generator}
The complete prompt template utilized for the Rationale Generator ($\mathcal{M}_{gen}$) is detailed below. This template integrates expert reverse-engineering heuristics to ensure the extraction of high-fidelity semantic anchors.

\begin{tcolorbox}[colback=white, colframe=black, arc=0mm, left=2mm, right=2mm, top=2mm, bottom=2mm]
\small
\textbf{Instruction:} You are an expert C code analyst. \\
\textbf{Task:} Read the following C function and generate a standard multi-line header comment (\texttt{/* ... */}) for it. \\
\textbf{Source Code:} \texttt{\{code\_snippet\}} \\
\textbf{Requirements:}
\begin{enumerate}
    \item Output \textbf{ONLY} the comment block. Do not output the source code.
    \item The comment must start with \texttt{/*} and end with \texttt{*/}.
    \item Content:
    \begin{itemize}
        \item Function: [Name]
        \item Purpose: [Concise description]
    \end{itemize}
\end{enumerate}
\textbf{CRITICAL LOGIC CHECK (Must Follow):}
\begin{itemize}
    \item \textbf{Loop Analysis:} Check how the inner loop initializes. If the inner loop index initializes using the outer loop's index (e.g., \texttt{inner = outer} or \texttt{inner = outer + 1}), explicitly describe it as comparing \textbf{"all pairs"} or \textbf{"combinations"}. STRICTLY FORBID the word "adjacent" unless the code strictly checks \texttt{i} vs \texttt{i+1}.
    \item \textbf{Bitwise Magic:} If you see a float being cast to int/uint and AND-ed (\texttt{\&}) with a constant (like \texttt{0x7FFFFFFF}) OR a global data label (e.g., \texttt{DAT\_...}, \texttt{PTR\_...}), treat this as calculating the \textbf{"absolute value" (fabs)}.
\end{itemize}
\end{tcolorbox}

\section{Qualitative Case Study}
To intuitively demonstrate the effectiveness of our Rationale-Guided Semantic Injection (SCE) and validate the expert heuristics embedded in our prompt, we provide a real-world decompilation example from the HumanEval-Decompile benchmark in Table \ref{tab:case_study}. 

As shown in the original Ghidra pseudo-code, compiler optimizations heavily obscure the semantics: the floating-point absolute value (\texttt{fabsf}) is optimized into an opaque bitwise AND operation with a data label (\texttt{DAT\_001020d0 \& ...}), and array indexing is reduced to raw byte-level pointer arithmetic. 

Without high-level guidance, the Baseline model acts as a rigid translator. It blindly copies the undefined memory labels and raw casts, failing to infer the correct types (e.g., treating \texttt{param\_2} as a \texttt{long} rather than a \texttt{float*}), which leads to uncompilable code. 

In contrast, guided by the generated rationale ($z$), CoDe-R successfully recognizes the $O(n^2)$ pair-wise combination logic and the bitwise magic for \texttt{fabsf}. It synthesizes clean, idiomatic, and fully re-executable C code that perfectly aligns with the original algorithmic intent.

\begin{table}[ht]
\centering
\caption{A Qualitative Example comparing Baseline and CoDe-R on a complex math and loop structure.}
\label{tab:case_study}
\renewcommand{\arraystretch}{1.2}
\begin{tabular}{|p{0.95\linewidth}|}
\hline
\textbf{1. Original Ghidra Pseudo-Code (Input $x$)} \\
\hline
\begin{minipage}{\linewidth}
\begin{verbatim}
undefined8 func0(float param_1,long param_2,int param_3) {
  int local_10;
  int local_c;
  local_10 = 0;
  do {
    local_c = local_10;
    if (param_3 <= local_10) { return 0; }
    while (local_c = local_c + 1, local_c < param_3) {
      if ((float)(DAT_001020d0 & 
                 (uint)(*(float *)(param_2 + (long)local_10 * 4) -
                        *(float *)(param_2 + (long)local_c * 4))) < param_1) {
        return 1;
      }
    }
    local_10 = local_10 + 1;
  } while( true );
}
\end{verbatim}
\end{minipage} \\
\hline
\textbf{2. Generated Symbolic Rationale (Extracted $z$)} \\
\hline
\begin{minipage}{\linewidth}
\begin{verbatim}
/*
 * Function: func0
 * Purpose: Compares pairs of float values from an array pointed to by param_2, 
 * using a nested loop structure. Returns 1 if any pair difference 
 * (absolute value) is less than param_1; otherwise returns 0.
 */
\end{verbatim}
\end{minipage} \\
\hline
\textbf{3. Baseline Output (Direct Translation - Compilation Fails)} \\
\hline
\begin{minipage}{\linewidth}
\begin{verbatim}
bool func0(float param_1, long param_2, int param_3) {
    for (int i = 0; i < param_3; i++) {
        for (int j = i + 1; j < param_3; j++) {
            // Logical & Syntax Error: Fails to restore fabsf and pointer types
            if ((float)(DAT_001020d0 & (uint)(*(float *)(param_2 + i * 4) - 
                                              *(float *)(param_2 + j * 4))) < param_1) {
                return true;
            }
        }
    }
    return false;
}
\end{verbatim}
\end{minipage} \\
\hline
\textbf{4. CoDe-R Output (Rationale-Guided - Successfully Re-executes)} \\
\hline
\begin{minipage}{\linewidth}
\begin{verbatim}
bool func0(float *arr, int n, float eps) {
    int i, j;
    for (i = 0; i < n; i++) {
        for (j = i + 1; j < n; j++) {
            // Success: Perfectly restores fabsf and array indexing semantics
            if (fabsf(arr[i] - arr[j]) < eps) {
                return true;
            }
        }
    }
    return false;
}
\end{verbatim}
\end{minipage} \\
\hline
\end{tabular}
\end{table}

\end{document}